# Breaking players' expectations:
# the role of non-player characters' coherence and consistency

Remi Poivet, Catherine Pelachaud, & Malika Auvray

**Abstract**—In video games, non-player characters (NPCs) have an essential role in shaping players' experiences. The design of their appearance and their behaviors can be manipulated in coherence and consistency to maintain players' expectations or on the contrary to induce surprise. The influence of NPCs' coherence and consistency on players' evaluation of them remains to be unveiled. To fill this gap, two experiments were conducted in the context of a military shooter game. Players' evaluation of NPCs' perceived intelligence and believability, were measured, as these two dimensions are fundamental for their adoption and engagement toward them. The first experiment investigated the impact of breaking players' initial expectations on the evaluation of NPCs. The second experiment focused on the influence of NPCs' coherence and consistency on both players' expectations and evaluation of NPCs, by means of a combination of questionnaires, behavioral, and physiological measures. Our results reveal that breaking players' expectations influence their evaluation of NPCs, with coherent and consistent design reinforcing expectations and incoherent design challenging them.

――――――――― ◆ ―――――――――

## 1 INTRODUCTION

IN video games, human players control a main character who interacts with non-player characters (NPCs) who populate the game. NPCs are designed with specific characteristics, chosen by game designers, that communicate their role in the narrative. NPCs can play different roles, such as allies or enemies. Game designers often use two characteristics to communicate NPCs' roles in the narrative: appearance and behaviors [1]. Players then form expectations about the role of NPCs based on the perceived coherence between their appearance and their behaviors [2]. Coherence refers to the correspondence between NPCs' features and players' stereotypes. It allows players to understand NPCs' behaviors within the wider context of the game world [3]. Players' expectations are reinforced by the consistency of NPCs' design throughout the game [4]. Consistency refers to the design parameters' alignment, such as appearance and behaviors, over time. It allows players to form expectations and ensure that NPCs' actions match predefined visual cues, in order to convey their role in the game [1]. For example, in a military shooter game, encountering an aggressive NPC who looks like a soldier influences the expectation of hostility when encountering a second NPC with a similar appearance. If the second soldier encountered confirms players' expectations, this will make it easier to predict hostility in subsequent encounters. The aim of maintaining both consistency and coherence between appearance and behaviors is to allow players to form accurate expectations of enemies or allies and, as a result, to improve the overall experience of the game [5]. In particular, when NPCs' behaviors match their appearance and fit logically into the game world, players can better immerse themselves in the narrative and make informed decisions throughout the game.

As part of their immersion, players form impressions of the NPCs based on their appearance and behaviors [6]. When interacting with NPCs, players assess their perceived intelligence and believability. Perceived intelligence is defined as the player's evaluation of NPCs' abilities to achieve their objective in the narrative, encompassing two dimensions: behaviors' understandability and performance [7]. Believability, on the other hand, is defined as the gap between players' expectations and their observations in the game, impacting their engagement in the narrative and their immersion [8]. It is crucial to assess the characteristics of perceived intelligence and believability, as they contribute to players' engagement and their appreciation of the overall experience of the game [1, 9]. To enhance the narrative experience, designers introduce twists and turns in the game [10]. This can be done by manipulating the coherence and consistency between the appearance of NPCs and their behaviors, to create surprise in the players' minds. For example, in the

• *Remi Poivet is with the Institut des Systèmes Intelligents et de Robotiques and Ubisoft, Paris, France. E-mail: remipoivet@yahoo.fr*
• *Catherine Pelachaud is with the CNRS-ISIR, Sorbonne Université, Paris, France. E-mail: catherine.pelachaud@isir.upmc.fr*
• *Malika Auvray is with the CNRS-ISIR, Sorbonne Université, Paris, France. E-mail: malika.auvray@sorbonne-universite.fr*

western action-adventure game Red Dead Redemption 2, the main character may encounter NPCs with a non-threatening appearance (e.g., a farmer), asking for help along the way. But, if the main character stops and helps them, they may suddenly become aggressive, violating players' initial expectations. This disconnect between appearance and behaviors does more than introduce an unexpected gameplay challenge, it triggers an emotional reaction in players, as they are likely to experience surprise, or even in the case of such an unexpected betrayal, frustration, fear, or anger. These experiences of strong incongruencies can lead to increased physiological arousal, which in turn improves immersion and enjoyment [11]. The intensity of these affective responses contributes to players' engagement with the narrative, reinforcing the dangerous and unpredictable nature of certain games, in our example, an American Western setting.

Stimulating players' affective states, for instance through incoherence between appearance and behaviors, is essential for maintaining players' engagement and immersion. On the other hand, coherence in NPC design helps players to form stable expectations and navigate the game world in a meaningful way. It's therefore difficult to strike a balance between these two requirements. In addition, the impact of violating players' expectation during their interactions with NPCS, both on players' physiological states and on their evaluations of NPCs' intelligence and believability, remains to be fully understood. We present two experiments in a military shooter game to study how players rely on their expectations and knowledge of NPCs to attribute intelligence and believability to them. The first experiment investigated the violation of players' expectations on their evaluation of NPCs' perceived intelligence and believability. The second experiment focused on the impact of consistency and the influence of coherence on players' expectations. To this end, we used behavioral measures and questionnaires assessing NPCs' perceived intelligence and believability. We also used physiological measures such as galvanic skin conductance, which is traditionally used in the exploration of several psychological processes linked to affective states such as variations in arousal and emotions, and heart rate, used to probe other affective states linked to stress. We formulate two hypotheses: (1) Breaking players' expectations will lower their evaluation of NPCs' perceived intelligence and believability. (2) After multiple interactions with NPCs that break players' initial expectations, participants update their expectations and ratings of perceived intelligence and believability.

The structure of the article includes a background section presenting NPCs' design factors, players' expectations, and the assessment of perceived intelligence and believability. Next, the rationale for the study is discussed and our methodology is described. The two experiments are respectively presented, along with their results. Each experiment is then discussed before delving into a general discussion. The article concludes by outlining limitations and suggesting future work.

## 2 BACKGROUND

### 2.1 Designing NPCs: the role of appearance and behaviors.

NPC creation involves defining consistent characteristics throughout the game experience [1]. For example, in shooter games, enemies are subdivided into archetypes defined by their appearances, the weapons they use, and their movement capabilities. As a result, game designers choose specific features to convey the desired game experience.

#### A. Appearance.

NPCs' appearance refers to their virtual embodiment in the game. The visual cues chosen to create NPCs' appearance aim to activate stereotypes in players' minds [2]. Based on these stereotypes, players can expect behaviors from NPCs such as understanding NPCs' social roles in the game [12]. For instance, in a military shooter game, an NPC dressed in military gear may be evaluated as more threatening than another NPC wearing casual clothes. In this scenario, the military gear triggers stereotypes linked to violence and implies potential aggressiveness towards the main character. Thus, as part of their identification of NPCs' roles, players anticipate NPCs' behaviors and abilities. In addition, the type of weapon carried by an enemy provides valuable clues as to their abilities [1]. For example, an NPC armed with a shotgun is likely to attempt close-range attacks, while one armed with a sniper rifle will engage from a distance. As a result, players can use these cues to make informed decisions during gameplay, such as whether or not to interact with an NPC. In addition, the context of the game plays a crucial role in interpreting NPCs' appearance. In a zombie shooter game, allies may have military gear, but players do not evaluate them as enemies due to the specific context of the game. Other factors, such as the shape or facial features of NPCs, can also be decisive in players' anticipation of their interactions. For instance, [13] investigated the evaluation of antagonists' appearance and morality based on the presence or absence of facial tattoos. This specific facial feature was found to be associated with immorality. In summary, appearance plays a crucial role in the experience of video game players, as it activates stereotypes in players' minds.

#### B. Behaviors.

NPCs' behaviors consist of predefined sequences of actions that shape how they interact with others. These behaviors can vary, from simulating casual activities to attacking the main character. However, the nature of their actions is constrained by their roles in the narrative, meaning that friendly NPCs are expected not to attack the main character, as it is not their intended purpose. As for enemies, the type of weapon they carry determines their behaviors, as illustrated in the previous comparison between shotguns and sniper rifles. Moreover, in shooter games, the aggressiveness of enemies can be manipulated by other parameters, such as their shooting accuracy or the damage they can inflict on the main character. Therefore, when creating NPCs, it's important to take into account the link between their behavior and their appearance, so that the design is consistent and coherent throughout the narrative.

## 2.2 Breaking players' expectations.

Over the course of their gaming experience, players predict new encounters based on NPCs' stereotypes and the match between their appearance and behaviors. To add complexity to the narrative and to surprise players, game designers can manipulate the design of NPCs to add unexpected interactions.

### A. How do we form and break expectations?

Players form enduring beliefs about NPCs' behaviors, known as expectations, through a cognitive process that involves gathering information from previous experiences, knowledge, and contextual cues within the game [14]. Thanks to their past video game experiences, players are familiar with the limitations and predefined behaviors of NPCs. Consequently, they expect NPCs to present stereotypical sequences of actions that confirm their roles [2]. When encountering enemies, players expect them to attack the main character until they are defeated. Over the course of the game experience, the main character may encounter several NPCs whose appearance and behavior are identical. The accumulation of similar interactions reinforces players' certainty in their decision-making. This mechanism relies on the human tendency to reinforce beliefs when events occur as expected, thus validating their expectations, even when the events occur randomly [4]. Violation of players' expectations refers to the introduction of unanticipated actions or events into the game experience [15] o break players' expectations, game designers can manipulate both the coherence and consistency of NPC design. On the one hand, incoherent NPC design would refer to the intentional mismatch between NPC appearance and behaviors. For example, an incoherent association between appearance and behaviors conveys false information to players' expectations. The violation of expectations triggers compensatory behaviors, which are reflected as exploratory behaviors in search of explanations [16]. For example in Ready or Not, aggressive NPCs with a civilian appearance can be explained by the context and the challenge of the game (i.e., identifying threats between NPCs is the internal challenge of the game). On the other hand, game designers can manipulate the consistency of NPC design, which means altering the occurrence of a certain association between appearance and behaviors (e.g., an NPC with a friendly appearance displays aggressive behavior) during a game experience.

### B. The impact of breaking expectations

Breaking expectations influences individuals' cognitive and emotional responses, as suggested by the ViolEx model [17]. When unexpected events violate people's expectations, it triggers a cognitive process of reevaluation and adjustment. The ViolEx model proposes three coping processes in response to expectation violations: accommodation, assimilation, and immunization, to maintain individuals' understanding of their surroundings. Accommodation involves adjusting beliefs after observing the outcome of unexpected events, while assimilation involves avoiding events that challenge initial expectations, and immunization involves reducing the relevance of the violation or reframing the meaning of the initial expectation. The selection of the coping strategy is influenced by individual factors such as personality traits, cognitive abilities, or past experiences [18, 19].

Breaking expectations can lead to cognitive flexibility and adaptability, as individuals adjust their beliefs based on new information [20], but it can also trigger emotional responses such as frustration or surprise [21]. Violation of expectations induces an appraisal process that may be moderated by the reward of the violation [22]. [23] measured the impact of inconsistent breaks of their expectations on participants' physiological responses. The experiment used a cyberball game where participants' characters had to throw a ball at NPCs. Researchers manipulated the possibility of the participants' avatar to throw the ball: either the participants are at the initiative of the action or, breaking participants' expectations, it is automatically done. The results indicated that breaking expectations triggered physiological responses similar to facing aversive events. Positive and negative attributions to the violations of expectations distinguish more or less favorable violations' outcomes [22].

Game designers may manipulate the design of NPCs to surprise players and convey positive violations of expectations, aiming to create twists, interest in the narrative, and increase the challenge of the game [24]. However, the impact of such manipulation on players' evaluation of NPCs remains unclear.

## 3 HOW TO MEASURE NPCS' PERCEIVED INTELLIGENCE AND BELIEVABILITY?

In video games, players' positive evaluations of NPCs' perceived intelligence and believability as defined in Section 1, are associated with narrative enjoyment, motivation to interact, and immersion [8, 25]. To evaluate NPCs' perceived intelligence, two dimensions are important: their understandability and performance [26]. Both dimensions are based on players' understanding of the NPCs' purposes and their efficiency in achieving their goals, as defined by their role in the narrative (e.g., ally or enemy). Regarding the influence of NPCs' role on expectations, for example, allies are designed to help the main character, while enemies are designed to confront them. Therefore, the enemies' perceived intelligence would depend on their abilities to defeat the main character. Factors such as enemies' strength or amount of health points increase the challenge of the game and thus, may affect players' evaluations of NPCs' perceived intelligence [9]. To investigate users' evaluation of artificial agents' perceived intelligence, [7] proposed an evaluation scale based on five semantic items: Incompetent/Competent, Ignorant/Knowledgeable, Irresponsible/Responsible, Unintelligent/Intelligent, Foolish/Sensible. The scores on each item are used to assess the two dimensions of understandability (Responsible, Intelligent, and Sensible scores) and performance (Competent and Knowledgeable scores). Using this scale, [27] observed a positive correlation between the evaluation of NPCs' aggressiveness and their performance scores, confirming the link between the perceived intelligence of enemies and the challenge they pose to the main character.

Aside from intelligence evaluation, players must believe in their interactions with NPCs to feel immersed and engaged in the narrative [8]. Because players evaluate NPCs based on their expectations of how they should behave, believable agents should closely match players' expectations during interactions [8]. Consequently, players' identification of NPCs' roles can influence their evaluation, as they will expect these agents to serve different purposes (e.g., interactions will differ when they occur with allies or enemies). Following [8]'s work, [28] proposed a believability scale composed of different dimensions that play a crucial role in players' evaluation of believable characters in narrative experiences. From [28]'s scale, players can indicate their evaluation of the believability of any NPCs (Where <X> is replaced by the evaluated NPC) along the dimensions:

- Awareness: < X > perceives the world around him/her.
- Behavior understandability: It is easy to understand what < X > is thinking about.
- Personality: < X > has a personality.
- Visual impact: < X >'s behavior draws my attention.
- Predictability: < X >'s behavior is predictable.
- Behavior coherence: < X >'s behavior is coherent.
- Change with experience: < X >'s behavior changes according to experience.
- Social: < X > interacts socially with other characters.

In our studies, we used the two scales of perceived intelligence [7] and believability [28] as a baseline to investigate how players perceive NPCs in terms of these attributes.

## 4 EXPERIMENTS

As seen above, players form expectations about NPCs based on the coherence between their appearance and behavior. Deliberately violating players' expectations by manipulating NPC design aims to create surprise and deepen the narrative. However, the effect of violating expectations on players' evaluation of NPCs remains to be investigated, which is our goal here.

During their immersion in the game, players evaluate NPCs' intelligence and believability based on their expectations. We hypothesize that decreasing the coherence of NPCs, by introducing a violation of expectations, will lead to a decrease in players' evaluation of NPCs' design and a decrease in their game experience. To investigate these hypotheses, two experiments were conducted in the context of a Ubisoft military shooter game. The first experiment investigated the influence of the coherence between NPCs' appearance and their behaviors on players' initial evaluation. The second experiment investigated the impact of consistency and coherence across multiple interactions in game sessions. Perceived intelligence and believability scales were used to measure players' evaluations of NPCs. Players' game experiences were assessed across multiple dimensions, including their in-game behaviors, physiological activity, and expectations of NPCs' behaviors based on their appearance.

## 5 GENERAL METHODS

### Procedure and materials

The two experiments were set in a military shooter game environment. The NPCs used in the experiments were selected from Ubisoft's video game Ghost Recon Breakpoint (2019). The first experiment was conducted online and participants watched four videos depicting interactions between a main character and NPCs with manipulated designs. The second experiment took place at Ubisoft Paris and involved video game sessions that preceded the viewing of the same videos that were used in Experiment 1. In both Experiments, participants rated their perceptions of the NPCs featured in the four videos using the Perceived Intelligence questionnaire [7] and the Believability questionnaire [28]. In addition, Experiment 2 investigated participants' interactions with NPCs and the evolution of their expectations based on NPC's appearance using a homemade questionnaire completed before, in between, and after the two game sessions, as well as through behavioral and physiological measures during the game sessions.

## Creation of NPCs' design

To create different NPC designs for the experiments, the appearance and behaviors of soldiers and civilians were manipulated to correspond to either a coherent or an incoherent association. In military shooter games, civilian-looking NPCs are typically associated with innocent characters. Their behaviors are characterized as submissive, for instance, they get down on their knees when the main character approaches them. Conversely, soldier-looking NPCs are expected to be enemies. Their behaviors are characterized as aggressive, for example they may shoot at the main character while running toward them. Thus aggressive soldiers and submissive civilians correspond to coherent associations (see Fig. 1) while submissive soldiers and aggressive civilians correspond to incoherent ones.

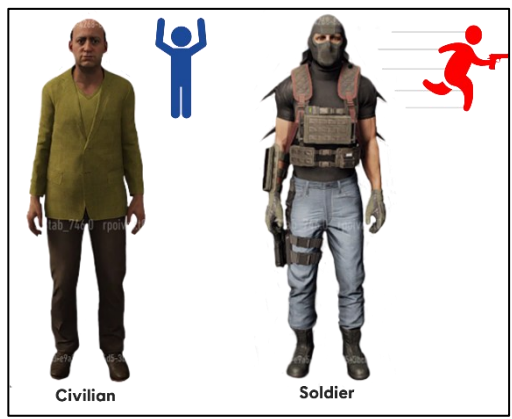

Fig. 1. NPCs' appearance and behaviors with a coherent association. On the left, the civilian's appearance is characterized by casual clothing and a non-muscular shape. Its behavior, graphically represented by the blue silhouette, corresponds to surrender when the main character approaches. On the right, the soldier's appearance is characterized by a muscular shape and military gear, with dark clothing and a covered face. Its behavior, illustrated by the red silhouette, corresponds to running towards the main character and shooting at them.

## Videos for participants' evaluation

Four videos were created, two with coherent and two with incoherent NPCs designs (see Fig. 2). In the videos, the camera follows the main character moving through a corridor with a NPC positioned at the end of it. To focus participants' attention on NPCs' appearance and behaviors, the videos have no sound. The video pauses after the interaction, with outcomes ranging from the main character being killed to the NPC surrendering. Here a coherent video may depict a soldier killing the NPC and an incoherent video may correspond to a soldier surrendering. The participants watched the videos once and were asked to rate NPCs' perceived intelligence and believability at the end of each video.

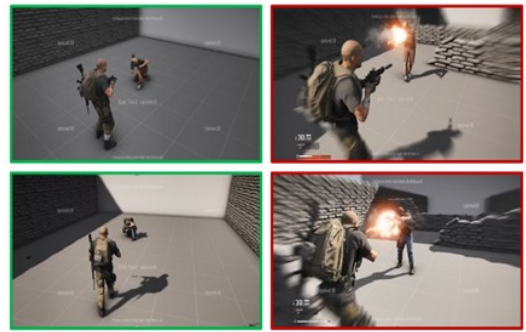

Fig 2. Screenshot of the videos illustrating the interactions between the main character (the one closest to the camera, seen from behind in all four images) and the four NPCs. The two images on the top show an NPC with a civilian appearance behaving non-aggressively (on the left) and aggressively (on the right). On the bottom, an NPC with a soldier appearance follows the same dichotomy. Note that watermarks are automatically generated for each image and are required by Ubisoft.

## 6 EXPERIMENT 1

### Participants

One hundred and two participants completed Experiment 1 (ninety-one men, ten women, and one other, mean age = 32.16 years, SD = 8.57). Participants were volunteer players registered on the Ubisoft User Research Laboratory platform (i.e., a mailing platform where players can voluntarily enter their information to participate in Ubisoft's research). The gender distribution can be attributed to the self-selection of participants who are already familiar with the military shooter game setting, potentially introducing a gender bias in their motivation for this specific type of game.

### Procedure and materials

All participants were contacted by email and were informed about the content of the research. The email informed participants that their participation was voluntary and that they could stop at any time. The experiment took approximately 15 minutes to complete and was conducted in accordance with [29]. Participants were instructed about the context of the game (i.e., a military shooter game) prior to viewing the videos. Participants watched the four videos in random order and rated the NPCs on the perceived intelligence and believability scales.

### Results and discussion

In order to maintain the representativeness of the dataset, no explicit outlier treatment was applied, hence all the collected data was included in the statistical analysis. A two-way ANOVA using the factors Appearance (soldier or civilian) and Coherence (coherent or incoherent design) was conducted on the participants' ratings of the scales' items. There was a main effect of the factor Appearance on the items "Visual impact" (df = 404, p = 0.001, η² = 0.023) and "Behavior coherence" (df = 404, p = 0.003, η² = 0.016) of the believability scale. Participants rated the behaviors of soldiers as drawing more attention compared to civilians' ones. However, civilians' behaviors were perceived as more coherent than soldiers' ones. There was a main effect of the factor "Coherence" on the items "Competent" (df = 404, p = 0.003, η² = 0.019), "Intelligent" (df = 404, p < .001, η² = 0.030), "Knowledgeable" (df = 404, p = 0.003, η² = 0.019), "Sensible" (df = 404, p < .001, η² = 0.111), and "Responsible" (df = 404, p < .001, η² = 0.056) from the perceived intelligence scale. Incoherent NPCs were perceived as less intelligent than coherent ones. There was a main effect of the factor "Coherence" on the items "Awareness" (df = 404, p = 0.024, η² = 0.012), "Behavior understandability" (df = 404, p < .001, η² = 0.093), "Predictability" (df = 404, p < .001, η² = 0.239), and "Behavior coherence" (df = 404, p < .001, η² = 0.222) from the believability scale. There was a significant interaction between the two factors on the items "Competent" (df = 404, p < .001, η² = 0.128), "Knowledgeable" (df = 404, p < .001, η² = 0.124), "Sensible" (df = 404, p < .001, η² = 0.033), "Responsible" (df = 404, p = 0.002, η² = 0.022) from the perceived intelligence scale and the items "Awareness" (df = 404, p < .001, η² = 0.044), "Personality" (df = 404, p = 0.011, η² = 0.016), "Visual impact" (df = 404, p < .001, η² = 0.085), "Predictability" (df = 404, p = 0.036, η² = 0.008), and "Behavior coherence" (df = 404, p = 0.048, η² = 0.007) from the believability scale.

Experiment 1 revealed a significant effect of the factors Appearance and Coherence on participants' ratings of perceived intelligence and believability (see Fig. 3). These results suggest that participants formed expectations about the behavior of NPCs based on the evaluation of their appearance and that their ratings were modified as a function of whether these expectations were confirmed (coherent association) or not (incoherent association). Incoherent NPCs were perceived as significantly less intelligent, with the strongest effects observed on the "Sensible" and "Responsible" items, suggesting that coherence plays a crucial role in participants' attributions of rational decision-making and accountability to NPCs. Similarly, believability ratings were found to be negatively impacted, with incoherent NPCs receiving lower scores on "Behavior coherence", "Predictability", "Behavior understandability", and "Awareness". These findings suggest that participants had a more positive evaluation of coherent NPCs, likely due to the alignment between their expectations and NPCs' behaviors. These findings are consistent with [8]'s definition of believability and [7]'s dimension of understandability of perceived intelligence. Participants observing incoherent NPCs were surprised and unable to accurately explain their behaviors, which led to a decrease in their ratings of NPCs' believability and perceived intelligence in all of the items, compared to the coherent ones (see Figure 3).

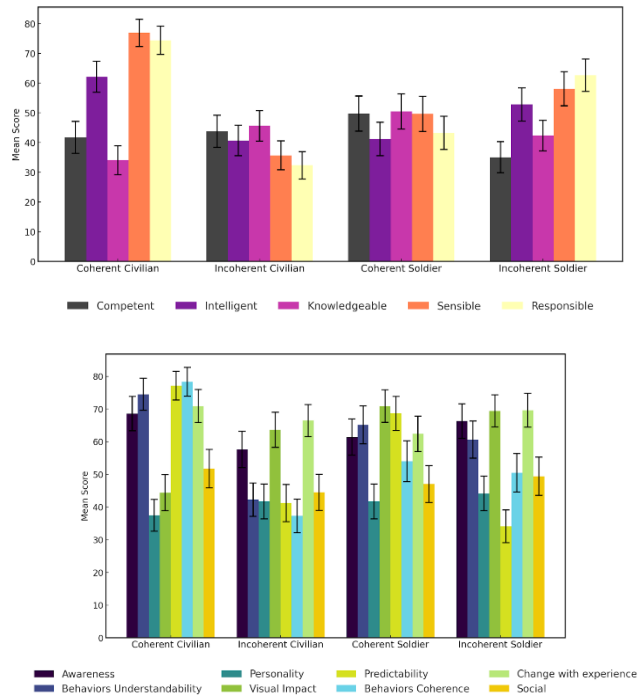

Fig. 3. Participants' mean ratings on the items from the perceived intelligence (on the top) and believability (on the bottom) scales, based on the coherence between appearance and behavior.

## 7 EXPERIMENT 2

### Participants

Ninety-one employees of Ubisoft (fifty-six men and thirty-five women, mean age = 27.5 years, SD = 6.29)

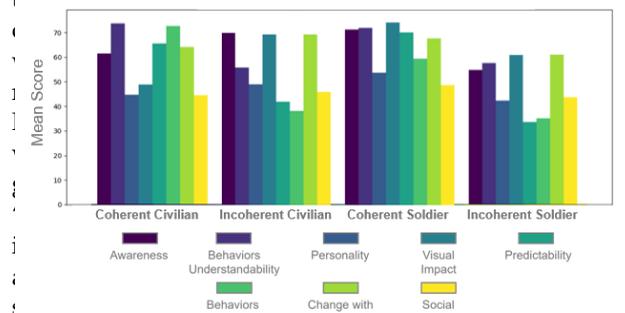

31.08; play often to play very often), while no difference was found in video game habits (mean = 76.57, SD = 24.05, play almost every day). All participants were contacted via email and provided with information regarding the general purpose of the research. The email highlighted the voluntary nature of participation, the various measures recorded, and the option to withdraw at any time. The experiment lasted approximately 50 minutes and was conducted in accordance with the principles outlined in [29].

**Procedure and materials**

Participants were randomly assigned to one of the three groups (see details in the following section) and they completed two game sessions during which behavioral and physiological measures were collected. Before the first, before the second, and after the second session, they were asked to complete a questionnaire about their intention to shoot at NPCs. Participants then watched four videos and were asked to rate the perceived intelligence and credibility of the NPCs after each video (similar to Experiment 1). Figure 4 illustrates the protocol, while information about the game sessions and the measures collected is detailed below.

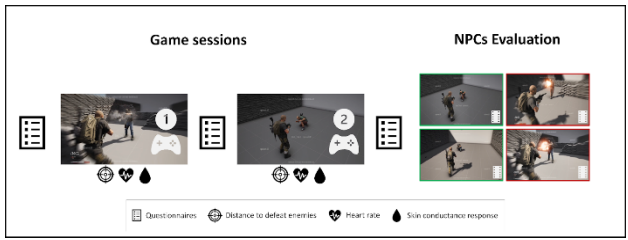

Fig. 4. Illustration of the procedure. Participants completed two military missions with questionnaires, behavioral, and physiological measures recorded during these sessions. Then, they watched 4 videos and completed a questionnaire on perceived intelligence and believability after each of them.

*Game sessions*

During each session of the game, the participants were engaged in a military mission. The primary objective for the main character, controlled by the participants, was to reach a specific location in the virtual environment. Along the way, ten NPCs were strategically placed to embody either innocents or enemies. The participants were instructed to shoot at those NPCs that they identified with high certainty as enemies. The first NPC encountered varied between game sessions, such that if the first NPC encountered was an enemy in session 1, it was an innocent in session 2, and vice versa.

The 3 groups of participants had specific NPCs' designs in their game sessions. Group 1 had a consistent and coherent NPC design, defined as encountering only aggressive soldiers (enemies) and submissive civilians (innocents). Group 2 had a consistent and incoherent NPC design where soldiers were submissive and civilians were aggressive. In group 3, the NPCs were inconsistent, alternating between coherent and incoherent designs, with soldiers and civilians displaying either aggressive or submissive behaviors. Fig 5 illustrates the design of the NPCs encountered by the 3 groups.

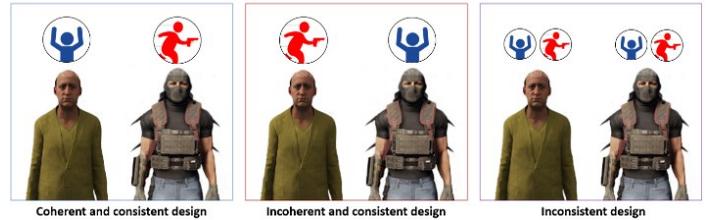

Fig. 5. Illustration of the three designs that were created for the game sessions. The red and blue figures illustrate the behaviors of the NPCs, either surrender (blue figure) or attack (red figure). On the left, Group 1: coherent design, civilians surrender when the main character approaches while soldiers try to defeat them. In the center, Group 2: incoherent design, the strict opposite of the coherent design. On the right, Group 3: inconsistent design, where both soldiers and civilians can display behaviors corresponding to surrender or attack.

*Questionnaires*

Participants were asked to indicate their intention to shoot at NPCs ("I would shoot at this NPC in the context of a military shooter game: Yes or No") based on their appearance (i.e., one appearance of a soldier and one of a civilian). Additionally, they were asked to rate the certainty of their intention to shoot on a scale ranging from "0 - I'm not certain at all" to "100- I'm totally certain". This questionnaire was completed before the first session, in between, and after the second game session. Behavioral measure

During the game sessions, participants facing enemies must defeat them to progress in the virtual environment. The distance between the main character and the defeated enemies was measured to reflect participants' decision-making pace in shooting enemies.

*Behavioral measure*

During the game sessions, participants facing enemies must defeat them to progress in the virtual environment. The distance between the main character and the defeated enemies was measured to reflect participants' decision-making pace in shooting enemies.

*Physiological measures*

During participants' interactions with NPCs, autonomous activities were recorded to assess their reactions in response to NPCs' behaviors. Skin conductance responses were measured, focusing on amplitude and rise time of the signal's peak for each interaction with NPCs. Skin conductance measurements were obtained from sensors placed on participants' left hand, specifically on the ring and little fingers. To ensure consistency, participants were instructed to use only their thumb, index, and middle fingers while using an Xbox controller. Additionally, their heart rate was recorded using Photoplethysmography (PPG). PPG data, capturing heart rate information, was collected by placing a sensor on the participants' left earlobe. To ensure accurate readings, participants were instructed to remove any earrings and precautions were taken to secure loose hair that might interfere with the sensor's contact with the skin. These measures were implemented to minimize potential sources of interference and enhance the reliability of the heart rate measurements throughout the study.

**Results and discussion**

In this experiment, participants completed three questionnaires regarding their decision to shoot at NPCs based on their appearance. They engaged in two game sessions and they rated four videos of an interaction between a main character and a NPC. This section will first investigate the results of the three questionnaires. Following that, the behavioral analysis of participants during their game sessions, followed by their SCR and heart rate, will be analyzed. Finally, the ratings provided by the participants for the four videos will be examined.

*Questionnaires' analysis*

In the questionnaires, participants indicated their decision to shoot at NPCs based on their appearance ("Would I shoot at this NPC? Yes or No") and they rated their certainty ranging from "0 - I'm totally unsure" to "100 - I'm totally sure"). We first analyzed the differences in the ratings of certainty before, in between, and after the game sessions. To do so, Chi-squared tests were conducted to analyze their intentions to shoot as a function of Appearance (soldier vs. civilian) and Group (groups 1, 2, or 3). Before the game sessions, there was no significant difference between groups regarding participants' intention to shoot at soldiers ($\chi2 = 1.042$, $p = 0.594$) or civilians ($\chi2 = 0.337$, $p = 0.845$). After the first game session, significant differences between the appearances of soldier ($\chi2 = 17.838$, $p < 0.001$) and civilian ($\chi2 = 8.128$, $p = 0.017$) were observed across all groups. After the second game session, these disparities persisted between soldier ($\chi2 = 25.082$, $p < 0.001$) and civilian ($\chi2 = 14.155$, $p < 0.001$) appearances. Second, we analyzed the evolution of notation across the questionnaires. To do so, a repeated-measure ANOVA was conducted to compare the effect of the factor Group on participants' certainty to shoot at soldiers and civilians (see Fig 6). For participants' certainty to shoot at soldiers, a Mauchly's test of sphericity on the repeated-measure ANOVA indicated that the assumption of sphericity had been violated ($\chi2(2) = 8.327$, $p < 0.001$). Thus, a Greenhouse-Geisser correction was applied, revealing a significant effect of Group (df = 3.665, $p < 0.001$, $\eta^2 = 0.063$). Post-hoc analysis using the Holm-Bonferroni method revealed significant differences between groups 1 and 3 ($p < 0.001$) and groups 2 and 3 ($p = 0.009$) after the two game sessions.

Participants from group 1, with the consistent and coherent NPCs' design were more certain of their decision to shoot at soldiers compared to participants from group 3, with an inconsistent NPCs' design. For participants' certainty to shoot at civilians, a Mauchly's test of sphericity on the repeated-measure ANOVA indicated that the assumption of sphericity had not been violated ($\chi2(2) = 2.551$, $p = 0.279$). The analysis revealed a significant effect of Group on participants' certainty to shoot at civilians (df = 4, $p = 0.017$, $\eta^2 = 0.033$). Post-hoc analysis using the Holm-Bonferroni method showed significant differences between groups 1 and 3 ($p < 0.001$) and groups 2 and 3 ($p = 0.004$) after the two game sessions. Participants in group 3 rated their certainty toward shooting or not at civilians lower compared to the other two groups after their game sessions.

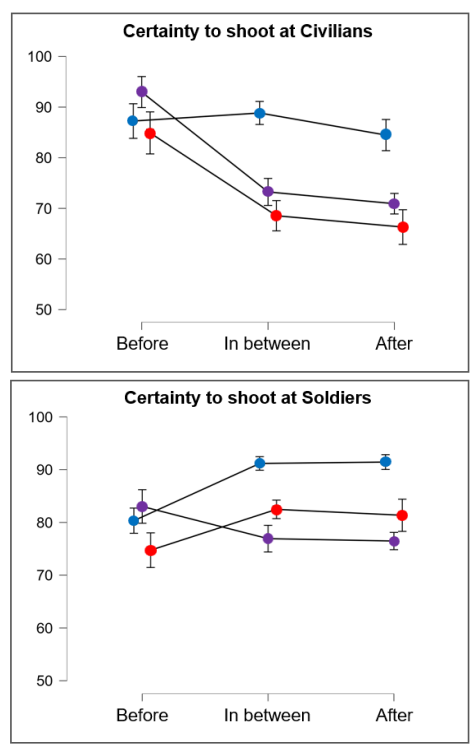

Fig. 6. Evolution of participants' certainty to shoot at NPCs with a civilian (top) and soldier (bottom) appearance over game sessions. Group 1 (coherent and consistent) is represented in blue, group 2 (incoherent and consistent) in red, and group 3 (incoherent and inconsistent) in purple.

*Behavioral analysis*

For the behavioral analysis, the distance (in meters inside the virtual environment) between the main character (controlled by the participants) and the defeated enemies was analyzed. A two-way ANOVA was performed to examine the effects of the factors Group and Session on the distance to defeat enemies (see Fig. 7). There was a main effect of the Group on the measured distance (df = 2, $p < 0.001$, $\eta^2 = 0.014$). Post-hoc analysis indicated that the distance between the main character and the enemies was larger in group 1 compared to group 2 ($p = 0.003$) and group 3 ($p = 0.003$).

On the other hand, there was no significant difference between groups 2 and 3 (p = 0.999). Furthermore, there was a main effect of Session on the distance to defeat enemies (df = 1, p < 0.001, η² = 0.016). The distance between the main character and the defeated enemies was significantly larger in the second session. There was no significant interaction between the factors Group and Session (df = 2, p = 0.411, η² = 0.002).

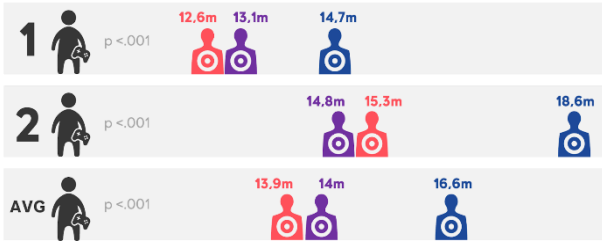

Fig. 7. Distance to defeat enemies in Session 1 and Session 2, for the 3 groups, group 1 (coherent and consistent) is represented in blue, group 2 (incoherent and consistent) in red, and group 3 (incoherent and inconsistent) in purple.

*Physiological analysis*

A two way ANOVA on participants' mean heart rate did not reveal any significant effect neither of Group (p= 0.345) nor of Session (p= 0.991). For participants' skin conductance responses, Two-way ANOVAs with the factors Group and Session were conducted on peaks' amplitude and peaks' rise time (see Fig. 8). There was a main effect of the factor Group on peaks' amplitude (df = 2, p < 0.001, η² = 0.018). Post-hoc analysis using the Holm-Bonferroni method indicated a significant difference between groups 2 and 3 (p < 0.001), but no significant difference between groups 1 and 2 (p = 0.053) or between groups 1 and 3 (p = 0.074). Participants in group 3 had significantly higher peaks during their interaction with NPCs compared to the other two groups. Additionally, there was a main effect of Session on the SCR's amplitude (df = 1, p = 0.015, η² = 0.006). Post-hoc analysis using the Holm-Bonferroni method showed a significant difference between the first and second sessions (p = 0.015), with all participants' peaks being lower in the second session. However, there was no significant interaction between the factors Group and Session on peaks' amplitude (df = 2, p = 0.578, η² = 0.001). Regarding peaks' rise time, there was a main effect of Group (df = 2, p = 0.005, η² = 0.010). Post-hoc analysis using the Holm-Bonferroni method revealed a significant difference between groups 1 and 2 (p = 0.004) and groups 2 and 3 (p = 0.044). However, there was no significant difference between groups 1 and 3 (p = 0.620). Participants in group 2 had peaks with significantly longer rise times compared to participants in the other two groups. There was no significant effect of Session (p = 0.799) nor any interaction between the two factors (p = 0.095) on peaks' rise time.

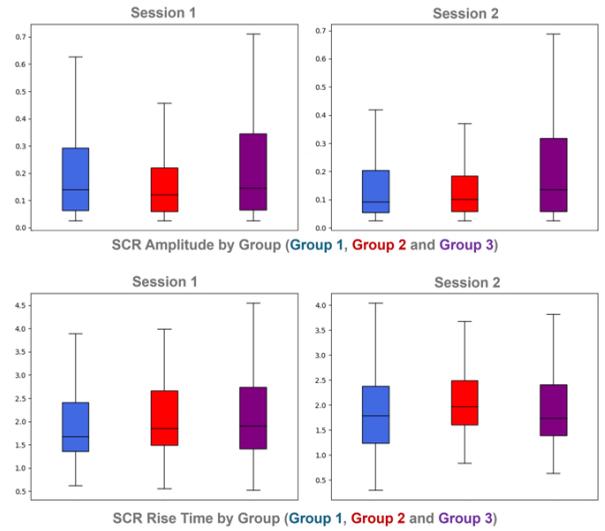

Fig. 8. Boxplots of participants' skin conductance response during their interaction with NPCs. Amplitude (top) and Rise time (bottom) of their response by session order.

*Evaluation of the videos*

After the game sessions, participants watched four videos and, for each one, they rated NPCs using the items of the perceived intelligence and believability scales. A three-way ANOVA was conducted on the items, with the factors Group, Appearance (soldier or civilian), and Coherence (coherent or incoherent design). There was no main effect of the factor Group on participants' ratings. There was a main effect of the factor Appearance on participants' rating. Soldiers were rated as more knowledgeable (df = 1, p = 0.001, η² = 0.024) and their behaviors were perceived as more attention-grabbing (df = 1, p = 0.002, η² = 0.023) than civilians, although civilians' behaviors were evaluated as more coherent (df = 1, p = 0.04, η² = 0.009). The factor Coherence had a main effect on several items: Coherent NPCs were perceived as more sensible (df = 1, p < 0.001, η² = 0.112), responsible (df = 1, p = 0.003, η² = 0.02), predictable (df = 1, p < 0.001, η² = 0.341), and understandable (df = 1, p < 0.001, η² = 0.114), while incoherent NPCs were perceived as having stronger personalities (df = 1, p = 0.034, η² = 0.012), more reactiveness (df = 1, p = 0.038, η² = 0.011), and higher visual impact (df = 1, p < 0.001, η² = 0.082). There was no significant interaction between the factors Group and Appearance. However, the interaction between Group and Coherence was significant for the items "Personality" (df = 2, p = 0.002, η² = 0.035), "Predictability" (df = 2, p < 0.001, η² = 0.060), "Behavior coherence" (df = 2, p = 0.002, η² = 0.026), and "Change with experience" (df = 2, p = 0.047, η² = 0.016) (see Fig. 9). Furthermore, there was a significant interaction between Appearance and Coherence for the items "Competence" (df = 1, p < 0.001, η² = 0.188), "Knowledgeable" (df = 1, p < 0.001, η² = 0.158), "Sensible" (df = 1, p < 0.001, η² = 0.114), "Responsible" (df = 1, p < 0.001, η² = 0.192), "Behavior understandability" (df = 1, p < 0.001, η² = 0.038), "Visual impact" (df = 1, p < 0.001, η² = 0.052), "Change with experience" (df = 1, p < 0.001, η²

= 0.03), and "Social" (df = 1, p = 0.031, η² = 0.013). Lastly, there was no significant triple interaction between the three factors on participants' ratings.

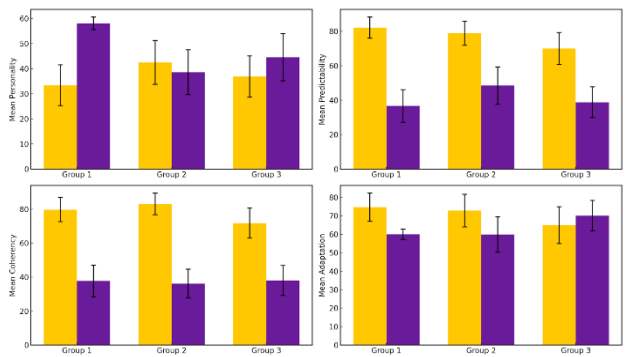

Fig. 9. Bar charts displaying the four items with significant differences influenced by the Group of game sessions and Coherence. Evaluation of coherent NPCs is depicted in yellow and incoherent ones, in purple. The items 'Behavior coherence' and 'Change with experience' from the believability scale have been renamed as 'Coherency' and 'Adaptation', respectively, for readability.

The results of Experiment 2 revealed that manipulating NPCs' design affects participants' interactions with NPCs, their expectations, and their evaluation of NPCs' perceived intelligence and believability. Participants' interactions and adjustments in expectations were investigated using a combination of behavioral, physiological, and subjective measures. The findings showed that participants had different behaviors and physiological reactions as a function of their assigned game design group, which were confirmed by their self-reported certainty to shoot at NPCs based on their appearance.

Behavioral data consisted of the distance to shoot at enemies during the game sessions, which varied across groups. Participants in group 1 maintained a greater distance to defeat enemies, likely due to the coherence and consistency of NPCs' design, as soldiers remained aggressive, confirming participants' expectations. In contrast, the other two groups had a shorter distance to defeat enemies, suggesting a longer decision time to shoot at enemies. However, participants in group 2 increased the distance between the two sessions, suggesting an adaptation of their expectations, i.e., taking into account the unusual association between civilian appearance and aggressiveness. In comparison, participants in group 3 kept the same distance between the two sessions, suggesting that the inconsistency of NPCs induced greater uncertainty in players and led them to expect more complex behaviors.

Regarding the measures of participants' skin conductance, participants in group 3 had a significantly higher amplitude in SCR than participants in groups 2 and 3. This suggests that they were more reactive toward NPCs during their game sessions. This finding aligns with behavioral data analysis, showing that participants in group 3 had a shorter distance to shoot at enemies, making them more prone to stress. The combination of the two analyses suggests that participants paid more attention toward NPCs in comparison to the other conditions. Regarding the measures on heart rate, the absence of a significant difference between groups can be attributed to participants' limited engagement in the game experience, which might not have been strong enough to elicit changes in their heart rates. For example, the absence of a compelling scenario or the simplicity of the virtual environment might have restrained participants' emotional investment during their interactions.

The questionnaires regarding the intention to shoot and its certainty revealed a significant difference between Group and Session. Prior to the game sessions, there were no significant differences between groups in terms of participants' intention to shoot at NPCs based on their appearance. However, after the game sessions, participants in group 1 were reinforced in their intention to shoot at soldiers, while participants in group 2 increased their intention to shoot at NPCs with a civilian appearance. In contrast, participants in group 3, who encountered inconsistent NPCs' designs, changed both their intention and certainty to shoot at NPCs across sessions. Their intention to shoot increased for NPCs with a civilian appearance and decreased for soldiers. At the same time, their certainty to shoot decreased across the sessions for both appearances, which shows an effect of inconsistency on their expectations about NPCs.

In summary, these results concerning players' interaction with NPCs suggest that consistency is an important parameter to enable players to form accurate and strong expectations about NPCs' roles. Hence, the manipulation of consistency confirms the Violex Model [17]: participants in group 2 adjusted their expectations based on the outcome of their interactions. Therefore, participants in groups 1 and 2 were able to make accurate decisions to shoot at NPCs based on their identification of NPCs' appearance. In contrast, participants in group 3 had to enhance their reactivity toward aggressiveness, leading to more stress during their game sessions and a reduced distance to defeat enemies.

Finally, participants' evaluations of NPCs in the videos provide insights into how their revised expectations affect perceived intelligence and believability. After the game sessions, participants in group 1 were the most affected by the unexpected behaviors of incoherent NPCs. They assign lower ratings on both scales because NPCs' behaviors contradicted their expectations, which were reinforced during the game sessions. However, participants rated higher specific items of the believability scale such as personality when confronted to unexpected NPCs. Conversely, participants in group 2 rated incoherent NPCs higher on the believability scale compared to the other groups. Specifically, their ratings for the understandability of incoherent NPCs' behavior were higher. From their game sessions, participants expected only aggressive civilians and submissive soldiers. However, similar to participants in group 1, their ratings for coherent NPCs were higher on the perceived intelligence scale and the other items of the believability scale, but this condition did not affect personality evaluation. Finally, participants in group 3 provided comparable ratings for both coherent and incoherent NPCs in the videos. In contrast to the participants in groups 1 and 2, the ratings of the incoherent NPCs did not improve as a result of the participants' interactions with them. Both coherent and incoherent NPCs received similar ratings on the item "Personality", unlike in group 1. Moreover, participants in group 3 assigned lower ratings to incoherent NPCs on the believability items when compared to Group 2, while coherent NPCs received ratings that were similar to those in groups 1 and 2. This suggests that the inconsistent condition led to lowered expectations among participants, influencing their evaluation of NPCs.

## 8 GENERAL DISCUSSION

The results from these two experiments, conducted in the context of a military shooter game, offer insights into the impact of breaking players' expectations on their in-game behaviors, associated physiological responses, ratings of NPCs' perceived intelligence and believability, and certainty to shoot. To break players' expectations, the coherence and consistency of NPCs' design (i.e., their appearance and behaviors) were manipulated.

The first experiment focused on breaking players' expectations by disrupting the coherence between appearance and behaviors, leading to inaccurate expectations. The second experiment examined the effect of manipulating both the coherence and the consistency between appearance and behaviors on players' expectations and gaming experience. Three designs were created to present coherent and consistent NPCs, incoherent and consistent NPCs, and inconsistent NPCs in the game sessions. In the following, we first discuss the results of Experiment 1 on coherence manipulation, and then the results of Experiment 2 on coherence and consistency manipulations.

### 8.1 Coherence manipulation (Experiment 1)

The first experiment aimed to investigate the effect of breaking players' expectations on their ratings of NPCs' perceived intelligence and believability. To this end, coherent and incoherent NPCs were created by manipulating their appearance and behaviors to convey inaccurate expectations. During the experiment, participants rated aggressive soldiers and submissive civilians higher in perceived intelligence and believability than submissive soldiers and aggressive civilians. This difference can be explained by the fact that players heavily rely on their expectations to evaluate NPCs. These expectations are influenced by their knowledge (i.e., the game setting) and stereotypes [2]. Here aggressive soldiers and submissive civilians are the most expected in a military game setting and are more in line with their purpose in the game. Encountering submissive soldiers, who deviate from this norm, may lead to misunderstandings and unexpectedness, thus reducing players' ratings of their perceived intelligence and believability. Players usually associate civilians with non-threatening roles, making it counterintuitive to engage with them as enemies. The decision to shoot at a civilian may be uncomfortable for players because it violates the moral compass they adopt while playing a military game [30]. Submissive civilians are evaluated more positively because they confirm players' expectations and help maintain their immersion in the game world. In summary, players tend to evaluate NPCs, be they soldiers or civilians, more positively when their behaviors confirm their expectations because it makes them easier to understand and aligns with their overall expectations of the game world.

### 8.2 Consistency manipulation (Experiment 2)

*Players' interaction with NPCs when manipulating coherence and consistency*

In Experiment 2, participants played a military shooter-type game with manipulated NPC designs, as described above. Participants were divided into three groups based on NPCs' design. Group 1 encountered NPCs with a coherent and consistent design, meaning that their behaviors were consistent with expectations based on military stereotypes. Group 2 encountered NPCs with an incoherent design, i.e. with behaviors that contradicted common expectations. Group 3 experienced an inconsistent design, where NPCs' behaviors varied, combining both coherent and incoherent designs. We investigated the extent to which NPCs' design affected players' expectations and their adaptation.

In Group 1, participants' expectations were reinforced, aligning with typical beliefs about enemies in military shooter games. Participants' high certainty ratings for shooting at NPCs based on their appearance were in line with the greater distance required to defeat enemies during game sessions. In Group 2, NPCs' design disrupted participants' initial expectations, akin to the findings in Experiment 1. Participants anticipated aggressive soldiers but encountered aggressive civilians, leading to shooting them. As mentioned in the Violex Model [17], repeated interactions with incoherent but consistent NPCs lead participants to adjust their expectations so that the fictional game still makes sense. This adaptation process can be seen in participants' SCR's rise time, which significatively exceeded that of the other two groups, suggesting increased attention [31, 32]. In addition, after two sessions, the adjustment of participants' expectations is reflected in an increase in their certainty scores when shooting at NPCs based on their appearance as well as in their increased distance to defeat enemies. In Group 3, participants encountered NPCs with an inconsistent design that completely disrupted their expectations. This was measured by declarative, behavioral, and physiological data. Specifically, inconsistency reduced participants' certainty that they would shoot at NPCs, resulting in a short distance to defeat enemies, and this occurred without any significant evolution across sessions. Participants in this group had to react to aggressiveness on the fly rather than anticipate it, in contrast to Groups 1 and 2. The resulting uncertainty induces a high SCR amplitude.

Furthermore, significant differences in physiological responses emerged between groups and sessions. Participants in groups 1 and 2 showed a marked decrease in the amplitude of their SCR across sessions, indicating their ability to accurately anticipate interactions with NPCs. However, participants in Group 3 had a different pattern of learning, remaining uncertain within and across sessions. Following [33] description of high amplitude in SCR, newly encountered NPCs in this group induced higher levels of stress in players, as they held the potential for fatal consequences regardless of their appearance.

Overall, manipulating NPCs' coherence and consistency influences the game experience. Consistency emerges as a crucial factor in managing players' expectations, while incoherency provokes unexpected events. Players adjusted their expectations when confronted with consistent but incoherent designs, which aligns with our initial hypotheses. On the other hand, inconsistency forced players to compensate for the decreased accuracy of their expectations by a constant increased reactivity to enemies' actions. This is reflected in their shorter distance to defeat enemies (suggesting longer decision times to shoot) and their increased stress levels, as evidenced by the higher amplitude of their skin conductance peaks. Game designers can use this finding to manipulate coherence and consistency, influencing both players' perceptions and their actions to create varied game experiences.

### Players' expectation of NPCs' intelligence and believability

Participants formed different expectations of civilians and soldiers during their gaming sessions. Ratings of the perceived intelligence and believability of NPCs in the videos were influenced by their appearance as participants rated soldiers as more attention-grabbing and competent than civilians. In addition, ratings differed as a function of NPC's coherence between appearance and behaviors and consistency (i.e., groups).

Participants in group 1 had their game sessions with coherent NPCs, and then when they watched the 4 videos, they were more affected by the unexpected behaviors of incoherent NPCs. They rated incoherent NPCs as more attention-grabbing and attributed more personality to them (positive expectancy violation). They also attributed less intelligence and believability to them than to their coherent counterparts (negative expectancy violation). Thus, unexpected behaviors lead to both positive and negative violations of expectations, likely because NPCs challenge players' understanding of the game while also piquing their interest. Following definition of believability [8], players' attribution of personality to NPCs may be of interest to game designers for creating immersive narratives and may serve as a powerful motivator for player engagement.

Participants in group 2 had their game sessions with incoherent but consistent NPCs. When they then watched the videos, they rated coherent NPCs as more believable and intelligent than incoherent ones. However, they still rated the believability of incoherent NPCs higher than participants in groups 1 and 3. This difference between groups can be explained along with [34] perspective, which defines believability as an evaluation based on players' expectations. Here, the initial expectations of aggressive soldiers and submissive civilians are not met and the correction of expectations led to a higher evaluation of incoherent NPCs than in Group 1, likely because aggressiveness leads to a higher evaluation of perceived intelligence and believability.

Participants in group 3 encountered inconsistent NPCs' design in their game sessions. They had similar ratings for coherent and incoherent NPCs. In particular, unlike participants in groups 1 and 2 they had similar ratings of the items of attention-grabbing and personality. Their constant uncertainty during the game experience may have led them to lower their expectations and not make a special effort to understand incoherent behaviors.

In summary, participants who encountered only coherent NPCs (group 1), reinforced their expectations and, as a result, exhibited heightened sensitivity to unexpected behaviors. This led to a reduction in their evaluation of perceived intelligence and believability but increased their attention and ratings of NPCs' personality in comparison to the other two groups. Participants who encountered only incoherent NPCs (Group 2) were able to form new expectations. Participants who encountered inconsistent NPCs (Group 3) were unable to form accurate expectations resulting in similar ratings for coherent and incoherent NPCs. Therefore, consistency in the appearance and behavior design of NPCs can increase players' investment of belief in NPCs. It is crucial for designers to be aware of the potential conflicts that inconsistency can create in players' immersion in the game. However, designers can still leverage inconsistency to create alternative game experiences. For example, the strategic placement of unexpected behaviors after multiple expected interactions can elicit surprise in players and enhance gameplay.

## 9 LIMITS AND FUTURE RESEARCH

There are two main limitations to this research: the selection of participants and the type of game used in the experiments. In Experiment 1, the participants were volunteer players registered in Ubisoft's user research database, which involves highly engaged players. Their high engagement may not be representative of all players' attitudes toward unexpected events in a game experience (i.e., incoherence and inconsistency). In addition, there was a predominance of young male participants, which may have introduced biases that affected some of the results (e.g., the strength of the physiological responses or the certainty of shooting). Reproducing the study with new selection criteria to include different levels of game expertise and more female players could provide a deeper understanding of the influence of violations of expectations on participants' perceptions as a function of individual characteristics (expertise and gender). In Experiment 2, participants were volunteers from Ubisoft. Although the game material was specifically adapted for the experiment and may not have reflected typical Ubisoft games, the fact that the experiment was conducted at Ubisoft, rather than in a more neutral environment may have influenced their evaluation of NPCs. In addition, the results obtained in our military shooter game context may not be generalizable to different contexts. In the context of our military game, players have expectations for NPCs to behave in a coherent manner. On the other hand, in science fiction settings, such as in the Fallout or The Last of Us video game series, stereotypes associated with post-apocalypse settings may modify players' expectations. In such a context, players are aware that they are likely to interact with NPCs exhibiting incoherent behaviors, which aligns with a presumed insane mind. As a result, the impact of incoherence and inconsistency on players' evaluation of believability and perceived intelligence differs in these contexts. Moreover, players' expectations may differ depending on the animation of NPCs. For example, in The Last of Us series, even if NPCs' behaviors are not more complex, their animation is much richer, particularly in terms of facial expressions. This can result in different players' expectations.

Consequently, future research endeavors should explore a wider range of games, including fantasy and cartoonish games, as well as settings with greater complexity, such as richer environments and more NPCs roles. In addition it would be interesting to diversify the participant base by considering factors such as gender, game preferences, and cultural background. This would provide a more thorough understanding of the factors that influence players' experiences and their expectations in different gaming contexts.

## 10 CONCLUSION

This study investigated the influence of NPCs' design (coherence and consistency) on players' expectations and ratings of perceived intelligence and believability. Experiment 1 revealed the consequences of violating players' expectations regarding NPCs' appearance and behaviors. Because players rely on their knowledge and prior experiences to form expectations, violating these expectations negatively affects their assessment of NPCs' abilities to achieve their purpose in the narrative. Experiment 2 focused on the impact of coherence and consistency of NPCs' design on players' game experience. Game experience was analyzed using a combination of questionnaires and behavioral and physiological data. First, the coherent and consistent design of NPCs reinforced players' expectations, leading to improved decision-making and ratings of NPCs' perceived intelligence and believability. Second, incoherent and consistent NPCs challenged players' expectations, but they were able to adjust their expectations over time. Third, inconsistent NPCs' design created uncertainty and prevented participants from forming accurate expectations, leading to increased stress as measured by the amplitude of their skin conductance responses. Overall, our study offers insights into the complex interplay between NPC design, player expectations, and their ratings of perceived intelligence and believability. These findings underscore the potential for game designers to create alternative game experiences by strategically manipulating players' expectations or by maintaining coherence and consistency in NPCs' design. In conclusion, a more profound comprehension of the impact of coherence and consistency on players' expectations can assist in the future design of more engaging and immersive player experiences.

## ACKNOWLEDGEMENTS

Remi Poivet is funded by the French ANRT and the Company Ubisoft.

## REFERENCES

[1] R. Giusti, K. Hullett, and J. Whitehead, "Weapon design patterns in shooter games," in Proc. 1st Workshop on Des. Patterns in Games, 2012, pp. 1-7, doi: 10.1145/2427116.2427119.

[2] H. Warpefelt, "Cues and insinuations: Indicating affordances of non-player character using visual indicators," in *Proc. DiGRA 2015: Diversity of play: Games - Cultures – Identities*, May. 2015. [Online]. Available: http://www.digra.org/digital-library/publications/cues-and-insinuations-indicating-affordances-of-non-player-character-using-visual-indicators/.

[3] A. Ortony, (2003). "On making believable emotional agents believable," in *Emotions in humans and artifacts*, Cambridge, MA, USA: MIT Press, 2003, ch. 6, pp. 189-211.

[4] D. Besnard, D. Greathead, and G. Baxter, "When mental models go wrong: Co-occurrences in dynamic, critical systems," Int. J. Human-Comput. Stud., vol. 60, no. 1, pp. 117-128, 2003, doi: 10.1016/j.ijhcs.2003.09.001.

[5] K. Cheng and P. A. Cairns, "Behaviour, realism and immersion in games," in *CHI'05 ext. abstr. on Human factors in comput. Systems, Apr. 2004*, pp. 1272-1275.

[6] A. Cafaro, H. H. Vilhjálmsson and T. Bickmore, "First impressions in human—agent virtual encounters," *ACM Trans. on Comput.-Human Interact. (TOCHI)*, vol. 23, no. 4, pp. 1-40, 2016, doi: 10.1145/2940325.

[7] C. Bartneck, D. Kulić, E. Croft, and S. Zoghbi, "Measurement Instruments for the Anthropomorphism, Animacy, Likeability, Perceived Intelligence, and Perceived Safety of Robots," *Int. J. of Social Robot.*, vol. 1, no. 1, pp. 71–81, 2009, doi: 10.1007/s12369-008-0001-3.

[8] A. B. Loyall, "Believable Agents: Building Interactive Personalities," Ph.D. dissertation, Carnegie Mellon University, 1997.

[9] C. Butcher and J. Griesemer, "The illusion of intelligence: The integration of AI and level design in Halo," in *Game Developers Conference*, San Jose, CA, March , 2002.

[10] E. Lavik, "Narrative structure in The Sixth Sense: a new twist in "twist movies"?," *The Velvet Light Trap*, vol. 58, no. 1, pp. 55-64, 2006, doi: 10.1353/vlt.2006.0031.

[11] N. Ravaja, et al. "Phasic Emotional Reactions to Video Game Events: A Psychophysiological Investigation," *Media Psychol.*, vol. 8, no. 4, pp. 353-367, 2006, doi: 10.1207/s1532785xmep0804_2.

[12] K. Rogers, M. Aufheimer, M. Weber, and L. Nacke, "Towards the Visual Design of Non-Player Characters for Narrative Roles," in Graphics Interface 2018, 2018, pp. 154–161, doi: 10.20380/GI2018.21.

[13] R. Pradantyo, M. V. Birk, and S. Bateman, "How the Visual Design of Video Game Antagonists Affects Perception of Morality," *Frontiers in Computer Science*, vol. 3, 2021, doi: 10.3389/fcomp.2021.531713.

[14] J. K. Burgoon and J. B. Walther, "Nonverbal expectancies and the evaluative consequences of violations," *Human Commun. Res.*, vol. 17, no. 2, pp. 232-265, 1990, doi: 10.1111/j.1468-2958.1990.tb00232.x.

[15] J. K. Burgoon, "Interpersonal expectations, expectancy violations, and emotional communication," *J. of lang. and soc. Psychol.*, vol. 12, no. 1-2, pp. 30-48, 1993, doi: 10.1177/0261927X9312100.

[16] J. Perez and L. Feigenson, "Violations of expectation trigger infants to search for explanations," *Cognition*, vol. 218, no. 104942, 2022, doi: 10.1016/j.cognition.2021.104942.

[17] C. Panitz, et al. " A Revised Framework for the Investigation of Expectation Update Versus Maintenance in the Context of Expectation Violations: The ViolEx 2.0 Model," *Front. Psychol.*, vol. 12, no. 726432, Nov. 2021, doi: 10.3389/fpsyg.2021.726432.

[18] M. Gollwitzer, T. Rothmund and P. Süssenbach, "The Sensitivity to Mean Intentions (SeMI) model: Basic assumptions, recent findings, and potential avenues for future research," Soc. Personal. Psychol. Compass, vol. 7, pp. 415–426, 2013, doi: 10.1111/spc3.12041.

[19] P. Süssenbach, M. Gollwitzer, L. Mieth, A. Buchner and R. Bell, "Trustworthy tricksters: Violating a negative social expectation affects source memory and person perception when fear of exploitation is high," *Frontiers in Psychol.*, vol. 7, no. 2037, pp. 1-12, 2016, doi: 10.3389/fpsyg.2016.02037.

[20] J. W. Sherman, A. Y. Lee, G. R. Bessenoff and L. A. Frost, "Stereotype efficiency reconsidered: Encoding flexibility under cognitive load," *J. Person. Soc. Psychol*. Vol. 75, pp. 589–606, 1998, doi: 10.1037/0022-3514.75.3.589.

[21] R. Casati and E. Pasquinelli,. "How can you be surprised? The case for volatile expectations," *Phenomenology and the Cogn. Sci.*, vol. 6, no. 1-2, pp. 171-183, 2007, doi: 10.1007/s11097-006-9028-9.

[22] J. K. Burgoon, "Expectancy violations theory," in *The int. encyclopedia of interpersonal commun.*, 2015. [Online]. Available: https://onlinelibrary.wiley.com/doi/full/10.1002/9781118540190.wbeic102

[23] M. Niedeggen, R. Kerschreiter and K. Schuck, "Loss of control as a violation of expectations: Testing the predictions of a common inconsistency compensation approach in an inclusionary cyberball game," *PLoS One*, 2019. [Online]. Available: https://journals.plos.org/plosone/article?id=10.1371/journal.pone.0221817

[24] M. Beigi, J. Callahan and C. Michaelson, "A critical plot twist: Changing characters and foreshadowing the future of organizational storytelling," *Int. J. of Manage. Rev.*, vol. 21, no. 4, pp. 447-465, 2019, doi: 10.1111/ijmr.12203.

[25] S. Moussawi, M. Koufaris and R. Benbunan-Fich, "How perceptions of intelligence and anthropomorphism affect adoption of personal intelligent agents," *Electron. Markets*, vol. 31, pp. 343-364, 2021, doi: 10.1007/s12525-020-00411-w.

[26] T. Koda and P. Maes, "Agents with faces: the effect of personification," in RO-MAN'96 – The 5th IEEE Int. Workshop Robot and Human Commun., 1996, pp. 189–194, doi: 10.1109/ROMAN.1996.568812.

[27] R. Poivet, A. de Lagarde, C. Pelachaud, and M. Auvray, "Evaluation of virtual agents' hostility in video games," IEEE *Trans. on Affect. Comput.* 2024, pp. 1-15, doi: 10.1109/TAFFC.2024.3390400.

[28] P. Gomes, A. Paiva, C. Martinho and A. Jhala, (2013). "Metrics for character believability in interactive narrative," in *Interactive Storytelling: 6th Int. Conf.*, 2013, pp. 223-228, doi: 10.1007/978-3-319-02756-2_27.

[29] World Medical Association, "WMA Declaration of Helsinki – Ethical Principles for Medical Research Involving Human Subjects," 2013. [Online]. Available:

https://www.wma.net/policies-post/wma-declaration-of-helsinki/

[30] B. DeVane and K. D. Squire, "The meaning of race and violence in Grand Theft Auto: San Andreas," *Games and Culture*, vol. 3, no. 3-4, pp. 264-285, 2008, doi: 10.1177/1555412008317.

[31] P. H. Venables, S. A. Gartshore, and P.W. O'riordan, "The function of skin conductance response recovery and rise time," *Biol. Psychol.*, vol. 10, no. 1, 1980, pp. 1-6, doi: 10.1016/0301-0511(80)90002-2.

[32] M. Shibagaki and T. Yamanaka, "Attention of preschool children: temporal variation of electrodermal activity during auditory stimulation," *Perceptual and motor skills*, vol. 72, no. 3, pp. 1115-1124, 1991, doi: 10.2466/pms.1991.72.3c.11.

[33] M. E., Dawson, A. M. Schell and D. L. Filion, "The electrodermal system," in *Handbook of psychophysiology*, Cambridge University Press, 2016, ch. 10, pp. 200-223.

[34] B. Magerko, (2007, November). "Measuring Dramatic Believability," in *AAAI Fall Symposium: Intelligent Narrative Technologies*, Nov. 2007. [Online]. Available: https://cdn.aaai.org/Symposia/Fall/2007/FS-07-05/FS07-05-014.pdf.

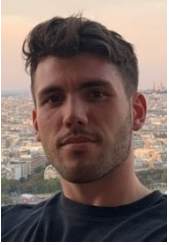
**Remi Poivet** is a former game design researcher from Ubisoft Paris and Ph.D student from ISIR, Sorbonne University. He received his PhD in Cognitive Science from Sorbonne University in 2023. His research interests are human evaluation of virtual agents and user experience in video games.

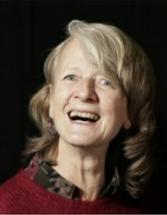
**Catherine Pelachaud** is CNRS Director of Research at ISIR, Sorbonne University. She received her Ph.D. in CS from University of Pennsylvania in 1991. Her research interest includes socially interactive agent, nonverbal communication (face, gaze, gesture and touch), and social interaction. With her research team, she has been developing an interactive virtual agent platform, Greta, that can display emotional and communicative behaviors. She is co-editor of the ACM handbook on socially interactive agents.

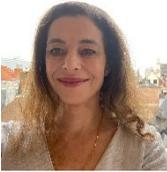
**Malika Auvray** is CNRS Director of Research at ISIR, Sorbonne University. She received her Ph.D. in Cognitive Sciences from EHESS, Paris in 2004. Her research interests include spatial and social cognition, multisensory perception and augmented cognition. She published more than 50 articles and book chapters and gave more than 300 communications on the topic. With her research team, she currently investigates the links between spatial and social perspective taking and ways to augment cognition by means of multisensory conversion.